\documentclass[prl,aps,10pt,longbibliography,superscriptaddress,twocolumn]{revtex4-2}
\usepackage{amsmath}
\usepackage{amssymb}
\usepackage{amsthm}
\usepackage{amsfonts}
\usepackage{enumerate}
\usepackage{latexsym}
\usepackage{bm}
\usepackage{graphicx}
\usepackage{subfigure}
\usepackage{color}
\usepackage[dvipsnames]{xcolor}
\usepackage[normalem]{ulem}
\usepackage[colorlinks]{hyperref}
\hypersetup{
  colorlinks,
  citecolor=Violet,
  linkcolor=Red,
  urlcolor=Blue}

\usepackage{cancel}
\usepackage{mathrsfs}
\usepackage{braket}

\newcommand{\beq}{\begin{equation}}
\newcommand{\eneq}{\end{equation}}
\input{epsf}

\begin{document}

\title{Physical identifiability of exceptional degeneracies}

\author{Zeng-Zhao Li}
\email{lizengzhao@iqasz.cn}
\affiliation{International Quantum Academy, Shenzhen, 518048, China}

\author{Xiao Xue}
\email{xiao.xue@ustc.edu.cn}
\affiliation{International Quantum Academy, Shenzhen, 518048, China}
\affiliation{Hefei National Laboratory, University of Science and Technology of China, Hefei 230088, China}

\begin{abstract}
Exceptional degeneracies are characterized by their Jordan structures, yet whether these structures can be physically identified depends on how the system is probed. Here we uncover an exact separation between dynamical accessibility and operational identifiability in an isospectral three-level system with fixed dissipation. Coherent control changes the Jordan partition from $(2,1)$ to $(3)$, but even a cyclic preparation accessing the entire emerging Jordan chain yields identical complete survival--loss outputs at a finite time, making the transition invisible to every final measurement. This blindness arises from destructive dynamical interference and can be overcome through preparation coherence: superpositions of individually blind states reveal hidden relative phases and produce an exactly linear, sign-sensitive population response, while their incoherent mixture remains blind. General Krylov-space criteria and distinct isospectral control paths further establish how physical access governs the detection and reconstruction of Jordan structures. Our results establish physical identifiability as a distinct aspect of exceptional-point physics, providing a framework for determining when hidden Jordan structures can be revealed through experimentally accessible quantum dynamics.
\end{abstract}

\date{\today}
\maketitle

{\it Introduction}---Exceptional points (EPs) are non-Hermitian degeneracies at which eigenvalues coalesce and eigenvectors become linearly dependent. Their defective eigenstructures underlie nonexponential dynamics, enhanced parameter responses, unconventional state transfer, and non-Hermitian quantum-information dynamics~\cite{Kato,AshidaGongUeda20AdvPhys,Bender2024,Heiss2012,ElGanainy2018,Miri2019,Bergholtz2021,Xue2026,Ding2022,Dembowski2001,Dembowski2004,Doppler2016,XuHarris16nature,Li2023,LiXue2026QuanAdv}, motivating spectral, dynamical, transport, and interference probes across physical platforms~\cite{Hodaei2017,Chen2017,Zhong2018,NaghilooMurch19nphys,WuRongDu19science,DingZhang21prl,WangJingChen21pra,Ryu2024,LiWhaley2024,LiXue2026}. Yet coincident eigenvalues do not specify the intrinsic defect structure: the sizes of the Jordan blocks determine the polynomial terms in the time-evolution operator, and the full Jordan partition provides a more complete characterization than exceptional-point order alone~\cite{LiWhaley2026}. 
Recent studies have addressed high-dimensional eigenspace coalescence~\cite{LiMa2026} and multiblock exceptional structures in open quantum systems~\cite{Shiralieva2026}. Operational approaches have examined preparation- and measurement-dependent visibility of Jordan contributions in quantum channels and time-resolved Krylov dynamics~\cite{Brandao2026,Ghosh2026}.  

A central challenge is to establish when the Jordan structures governing non-Hermitian dynamics can be inferred from physical observations. This raises a counterintuitive question: can a change in Jordan structure remain invisible in the complete quantum output even when the preparation accesses the entire emerging Jordan chain? Addressing this question requires separating structural changes from spectral motion and dissipation, while identifying the physical resources needed to recover hidden information. Neither spectral sensitivity~\cite{Lau2018} nor dynamical accessibility alone guarantees structural identifiability. Within the broader context of quantum system identification~\cite{Burgarth2012,Burgarth2014}, we focus on the injectivity of a fixed physical probing protocol rather than full process reconstruction.

In this Letter, we demonstrate an exact separation between dynamical accessibility and physical identifiability in an isospectral three-level system with fixed dissipation, where coherent control changes the Jordan partition from $(2,1)$ to $(3)$. Although a fixed preparation becomes cyclic and accesses the entire emerging Jordan chain, its complete survival--loss output coincides with that of the reference system at a finite interrogation time, making the transition invisible to every final measurement.

The blindness arises from destructive dynamical interference, not incomplete Jordan-chain access. Crucially, preparation coherence recovers hidden information: superpositions of individually blind states reveal relative dynamical phases invisible in their incoherent mixture, enabling an exactly linear, sign-sensitive response in a fixed population measurement. 
Beyond this example, we establish general criteria distinguishing isolated from persistent blindness and identify sufficient physical resources for detecting and reconstructing Jordan structures. 
We further show that distinct coherent-control paths can produce different operational responses despite identical spectra, dissipation, and Jordan transitions. Thus, physical identifiability is not determined by Jordan structure alone, but by how that structure is accessed and probed.

\begin{figure}[t]
\centering
 \includegraphics[width=0.96\columnwidth]{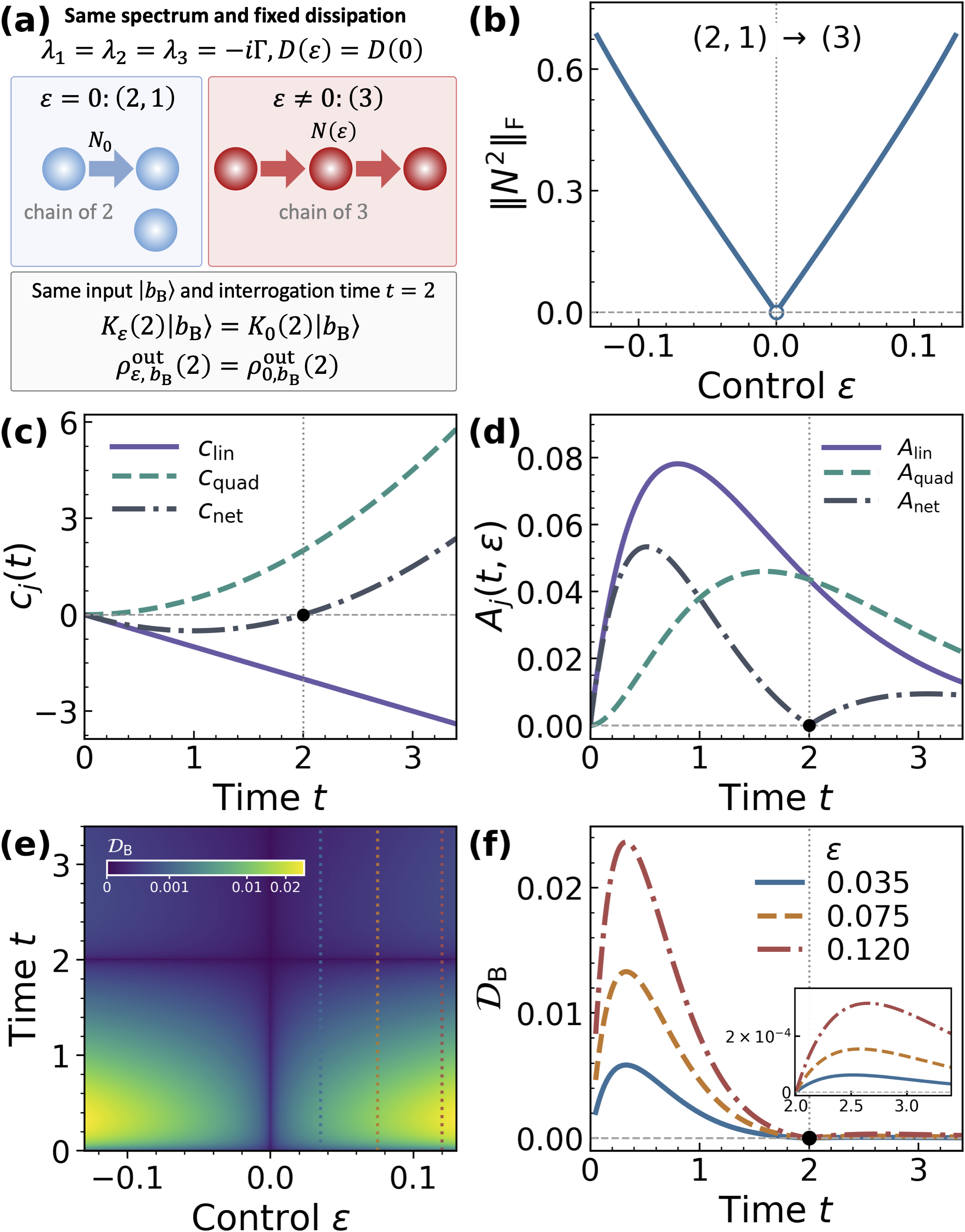} 
\caption{
\textbf{Jordan-structure transition, destructive interference,
and exact finite-time dynamical blindness.}
(a) Schematic of the exactly isospectral, fixed-dissipation family. A nonzero coherent control changes the Jordan partition from $(2,1)$ to $(3)$, yet the reference and controlled Hamiltonians yield identical complete survival--loss outputs for the same preparation $|b_{\rm B}\rangle$ at $t=2$.
(b) Structural diagnostic $\|N(\varepsilon)^2\|_{\rm F}$, where $\|\cdot\|_{\rm F}$ denotes the Frobenius norm, along the isospectral branch. Its vanishing at $\varepsilon=0$ distinguishes the $(2,1)$ partition from $(3)$ at nonzero $\varepsilon$. 
(c) Signed coefficients $c_{\rm lin}=-t$, $c_{\rm quad}=t^2/2$, and $c_{\rm net}=c_{\rm lin}+c_{\rm quad}$ of the linear, quadratic, and net contributions to the surviving-amplitude difference. The two nonzero contributions cancel at $t=2$.
(d) Corresponding surviving-amplitude magnitudes $A_j(t, \varepsilon) =e^{-\Gamma t}|c_j(t)|\|w_\varepsilon\|$ ($j={\rm lin,quad,net}$), for $\varepsilon=0.10$. The net amplitude vanishes at $t=2$ despite nonzero individual contributions.
(e) Complete survival--loss trace distance $\mathcal D_{\rm B}(t,\varepsilon)$ for the fixed preparation $|b_{\rm B}\rangle$. The dark horizontal band at $t=2$ indicates exact dynamical blindness, while the vertical dotted lines mark the control values used in (f).
(f) Time-dependent complete-output trace distance $\mathcal D_{\rm B}(t,\varepsilon)$ for $\varepsilon=0.035$, $0.075$, and $0.120$. All three curves vanish exactly at $t=2$, as marked by the black dot. The inset magnifies the nonzero distinguishability reappearing after $t=2$. The common loss offset is $\Gamma=1.25$.
}

\label{fig:fig1}
\end{figure}

{\it Exact invisibility of a Jordan transition}---We consider a three-level open quantum system governed by the effective non-Hermitian Hamiltonian $H(\varepsilon)= C(\varepsilon)-iD$, with $D=\Gamma I+F$ and $F= \operatorname{diag}(1,0,-1)$, where $C(\varepsilon)=C^\dagger(\varepsilon)$ represents coherent controls and $\Gamma>1$ ensures strictly positive, level-dependent loss rates. The parameter \(\varepsilon\) modifies only the Hermitian part of the Hamiltonian, leaving the dissipative operator \(D\) unchanged. 
We describe the surviving sector by the no-jump propagator $K_\varepsilon(t)=e^{-iH(\varepsilon)t}$ and retain the unnormalized surviving state together with an orthogonal loss state $|L\rangle$ carrying the total loss probability. This survival--loss output is unconditional, requiring no postselection, although individual loss channels and time-resolved jump records are not resolved.

In contrast to conventional spectral unfolding, we construct an exactly isospectral control path that changes the Jordan structure without moving any eigenvalue or modifying the dissipation. Writing \(H(\varepsilon)=-i\Gamma I+N(\varepsilon)\), we choose, in the basis $\{|1\rangle,|2\rangle,|3\rangle\}$ 
\begin{equation}
N(\varepsilon)=
\begin{pmatrix}
-\varepsilon-i&\varepsilon+iy&1-y+2i\varepsilon\\
\varepsilon-iy&2\varepsilon&-y+i\varepsilon\\
1-y-2i\varepsilon&-y-i\varepsilon&-\varepsilon+i
\end{pmatrix},
\label{eq:N}
\end{equation}
where $y(\varepsilon)=(1-\sqrt{1-27\varepsilon^2})/3$ and $|\varepsilon|<1/\sqrt{27}$. 
The identity $9\varepsilon^2+3y^2-2y=0$ gives $N^3(\varepsilon) =0$, fixing all three eigenvalues of $H(\varepsilon)$ at $-i\Gamma$. At $\varepsilon=0$, $N_0^2=0$ and $N_0\neq0$, yielding the Jordan partition $(2,1)$. For sufficiently small nonzero $\varepsilon$, $N^2(\varepsilon)\neq0$, giving the partition $(3)$. The transition occurs already at first order in the Hermitian control: $V=\partial_\varepsilon H|_0\neq0$ and $\{N_0,V\} =\partial_\varepsilon N^2|_0\neq0$ [Fig.~\ref{fig:fig1}(a,b)]. This three-level realization is minimal among isospectral Hamiltonians with identical dissipation~\cite{SM}.

Despite this genuine structural transition, its dynamical signature can disappear completely for a physically accessible preparation. Consider the preparation $|b_{\mathrm B}\rangle=(-3i|1\rangle+i|2\rangle+ |3\rangle)/\sqrt{11}$ and the fixed interrogation time $t=2$. Because $N^3(\varepsilon)=0$, the propagator terminates exactly at second order, $e^{-2i N(\varepsilon)} =I-2i N(\varepsilon) -2N^2(\varepsilon)$. Direct evaluation gives $e^{-2iN(\varepsilon)}|b_{\mathrm B}\rangle =(i|1\rangle+i|2\rangle -3|3\rangle)/\sqrt{11}$, independently of $\varepsilon$. Thus the surviving output amplitude, including its norm, is exactly unchanged throughout the analytic Jordan-transition branch.

The cancellation can be characterized at arbitrary interrogation times. Defining $\Delta N(\varepsilon) =N(\varepsilon) -N_0$, we find the exact identity $N^2(\varepsilon) |b_{\mathrm B}\rangle =-i\Delta N(\varepsilon)|b_{\mathrm B}\rangle$, which yields
\begin{equation}
\left[e^{-itN(\varepsilon)}-e^{-itN_0}\right]
|b_{\mathrm B}\rangle
=\frac{it(t-2)}{2}
\Delta N(\varepsilon)|b_{\mathrm B}\rangle.
\label{eq:finite-time-cancellation}
\end{equation}
To resolve the interference mechanism, define $|w_\varepsilon\rangle =i\Delta N(\varepsilon) |b_{\rm B}\rangle$. Equation~(\ref{eq:finite-time-cancellation}) then gives the surviving-amplitude difference $|\Delta\psi_{\rm surv}\rangle \equiv(K_\varepsilon-K_0) |b_{\rm B}\rangle =e^{-\Gamma t}(c_{\rm lin}+c_{\rm quad}) |w_\varepsilon\rangle$, where $c_{\rm lin}=-t$ and $c_{\rm quad}=t^2/2$ are the real coefficients of the linear and quadratic propagator contributions, respectively.
Their sum $c_{\rm net}=t(t-2)/2$ vanishes at $t=2$, although both contributions are nonzero [Fig.~\ref{fig:fig1}(c)]. The corresponding surviving-amplitude magnitudes, $A_j=e^{-\Gamma t}|c_j|\|w_\varepsilon\|$ ($j={\rm lin,quad,net}$), are shown in Fig.~\ref{fig:fig1}(d). Thus, the exact blind spot results from destructive interference between nonzero dynamical contributions, rather than from vanishing Hamiltonian sensitivity.

Remarkably, the blindness persists despite complete dynamical accessibility. The dimension of the Krylov subspace $\mathcal K_b =\operatorname{span}\{H^n|b\rangle:n\geq0\}$ characterizes the dynamical accessibility of a preparation $|b\rangle$. For $|b_{\rm B}\rangle$, this dimension increases from two at $\varepsilon=0$ to three for nonzero $\varepsilon$, making the preparation cyclic after the transition. 
Unlike restricted-access or readout-dependent invisibility~\cite{Brandao2026,Ghosh2026}, this blind spot occurs despite cyclic preparation and holds for the complete survival--loss output~\cite{SM}.

To establish operational indistinguishability, we consider the complete flagged output
\begin{equation}
\rho_{\varepsilon,b}^{\mathrm{out}}(t)
= 
K_\varepsilon(t)|b\rangle\langle b|K_\varepsilon^\dagger(t)
\oplus
\bigl[1-\|K_\varepsilon(t)b\|^2\bigr]|L\rangle\langle L|.
\end{equation}
For $|b_{\mathrm B}\rangle$ at $t=2$, the exact equality of the surviving amplitudes implies $\rho_{\varepsilon, b_{\mathrm B}}^{\mathrm{out}}(2) =\rho_{0, b_{\mathrm B}}^{\mathrm{out}}(2)$, as illustrated in Fig.~\ref{fig:fig1}(a).
Consequently, no measurement on these flagged outputs, even a collective measurement on arbitrarily many independent copies, can distinguish the reference point from a nonzero control value using this fixed preparation and interrogation time. This is a protocol-specific blind spot, not an absence of structural change.

We quantify operational distinguishability by the complete-output trace distance~\cite{Helstrom1976} $\mathcal D_b(t,\varepsilon) =\frac12 \|\rho_{\varepsilon,b}^{\mathrm{out}}(t) -\rho_{0,b}^{\mathrm{out}}(t)\|_1$, where $\|\cdot\|_1$ denotes the trace norm. For $|b\rangle=|b_{\mathrm B}\rangle$, we denote it by $\mathcal D_{\mathrm B}(t,\varepsilon)$. 
The distinguishability landscape reveals a continuous line of exact blindness at $t=2$ throughout the isospectral branch [Fig.~\ref{fig:fig1}(e)], while representative time traces resolve the recovery of distinguishability away from this blind time [Fig.~\ref{fig:fig1}(f)].
The common loss offset affects only the signal magnitude: since $K_\varepsilon(t) =e^{-\Gamma t} e^{-itN(\varepsilon)}$, the complete-output trace distance factorizes as $\mathcal D_b(t,\varepsilon) = e^{-2\Gamma t} \mathcal F_b(t,\varepsilon)$, where $\mathcal F_b$ is independent of $\Gamma$. Thus, varying $\Gamma>1$ changes the distinguishability but neither shifts nor removes exact blind spots~\cite{SM}.
Detuning the interrogation time by $\delta t$ restores distinguishability, with the local response $\mathcal D_{\mathrm B}(2+\delta t, \varepsilon) \propto |\varepsilon\delta t|$. The corresponding root-fidelity analysis gives an optimal sample complexity $n_{\mathrm{opt}} =\Theta([\varepsilon^2 (\delta t)^2]^{-1})$ for discriminating the two outputs at a fixed target error probability (see Supplemental Material~\cite{SM}).

\begin{figure}[t]
\centering
 \includegraphics[width=0.99\columnwidth]{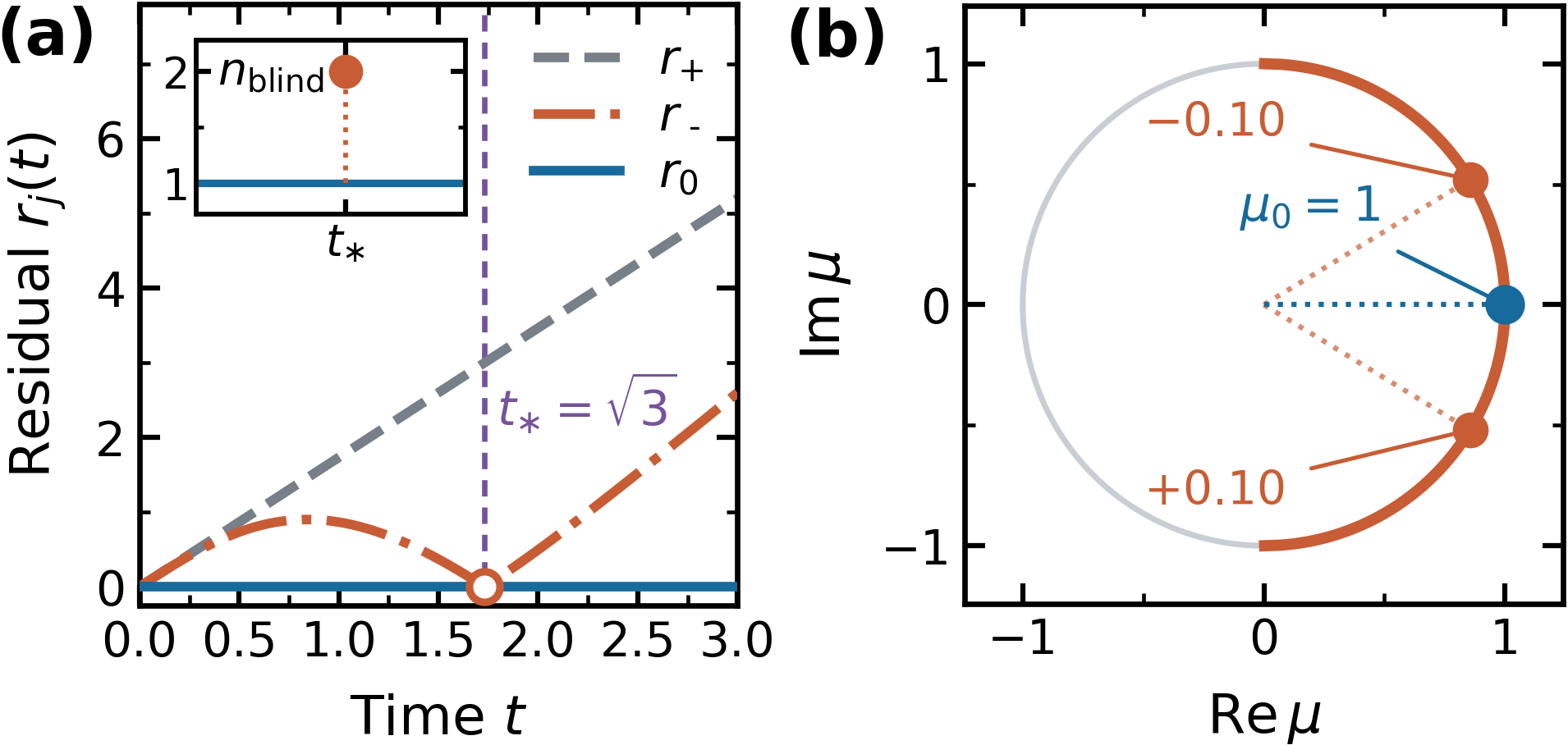} 
\caption{\textbf{Classification of blind preparations and hidden dynamical phases.}
(a) Common-eigenvector residuals $r_j(t)=\|(I-P_j)B(t)v_j(t)\|/\|v_j(t)\|$ for eigenvectors $v_j(t)$ of $A(t)$, where $P_j=v_jv_j^\dagger/(v_j^\dagger v_j)$ and $j=0,-,+$. A zero residual identifies a common-eigenvector candidate; exact blindness additionally requires a unit-modulus relative-propagator eigenvalue. The residual $r_0$ vanishes at all positive times, whereas $r_-$ vanishes only at $t_*=\sqrt{3}$ and $r_+$ has no positive-time zero. The inset shows the number $n_{\rm blind}$ of linearly independent blind directions fixed along the entire analytic control branch.
(b) Relative-propagator eigenvalues at $t_*$: $\mu_0=1$ and $\mu_-(\varepsilon)=1-3y(\varepsilon) -3i\sqrt{3}\,\varepsilon$. Both have unit modulus, while $\mu_-$ acquires a nontrivial phase relative to $\mu_0$. The highlighted points correspond to $\varepsilon=\pm0.10$.
}

\label{fig:fig2}
\end{figure}

{\it Hidden-phase structure of blind preparations}---To classify preparations that remain exactly blind throughout the isospectral control branch at a fixed interrogation time, we introduce the relative propagator $R_\varepsilon(t)= K_0(t)^{-1} K_\varepsilon(t) = e^{itN_0} e^{-itN(\varepsilon)}$. A preparation $|b\rangle$ is exactly blind if and only if $R_\varepsilon(t) |b\rangle = \mu_\varepsilon |b\rangle$ with $|\mu_\varepsilon|=1$: the two surviving amplitudes then differ only by an unobservable global phase.

The relative propagator has the exact decomposition $R_\varepsilon(t) =I+\varepsilon A(t) +y(\varepsilon)B(t)$, where $A(t)$ and $B(t)$ are independent of $\varepsilon$~\cite{SM}. Preparations fixed along the entire control branch must therefore be common eigenvectors of $A(t)$ and $B(t)$, with the corresponding relative-propagator eigenvalue of unit modulus. The common-eigenvector residuals in Fig.~\ref{fig:fig2}(a) reveal one blind direction at every positive time and a second at the unique time $t_*=\sqrt3$.

The blind direction existing at every positive time is~\cite{SM}
\begin{eqnarray}
|b_0(t)\rangle &=& \frac{-i(t+1)|1\rangle+i|2\rangle +(t-1)|3\rangle}{\sqrt{2t^2+3}}, \\
\,
R_\varepsilon(t)|b_0(t)\rangle &=&\mu_0|b_0(t)\rangle,
\qquad \mu_0=1.
\label{eq:blind-family}
\end{eqnarray}
In particular, $|b_0(2)\rangle=|b_{\mathrm B}\rangle$. Thus the $t=2$ cancellation belongs to a continuous family of time-dependent blind preparations, rather than being an isolated algebraic coincidence.

At the special time $t_*=\sqrt3$, a second blind direction, linearly independent of $|b_0(t_*)\rangle$ and fixed along the control path, is~\cite{SM}
\begin{align}
|b_-\rangle &= \frac{-i(1+\sqrt3)|1\rangle-2i|2\rangle
+(\sqrt3-1)|3\rangle}{\sqrt{12}}, \\
R_\varepsilon(t_*)|b_-\rangle
&=\mu_-(\varepsilon)|b_-\rangle,
\label{eq:second-blind-direction}
\end{align}
where $\mu_-(\varepsilon) =1-3y(\varepsilon) -3i\sqrt3\, \varepsilon$ has unit modulus throughout the control branch. Unlike $\mu_0=1$, it acquires a nontrivial phase for $\varepsilon\neq0$ [Fig.~\ref{fig:fig2}(b)]. This relative phase is invisible to either preparation individually but becomes accessible through coherent superposition.

This classification concerns $\varepsilon$-independent preparations that remain blind throughout the analytic control branch, without excluding parameter-dependent blind directions at other times (see Supplemental Material~\cite{SM} for the derivation and uniqueness proof). The coexistence of two blind directions with distinct phase factors provides the mechanism for coherence-enabled recovery.

\begin{figure}[t]
\centering
 \includegraphics[width=0.99\columnwidth]{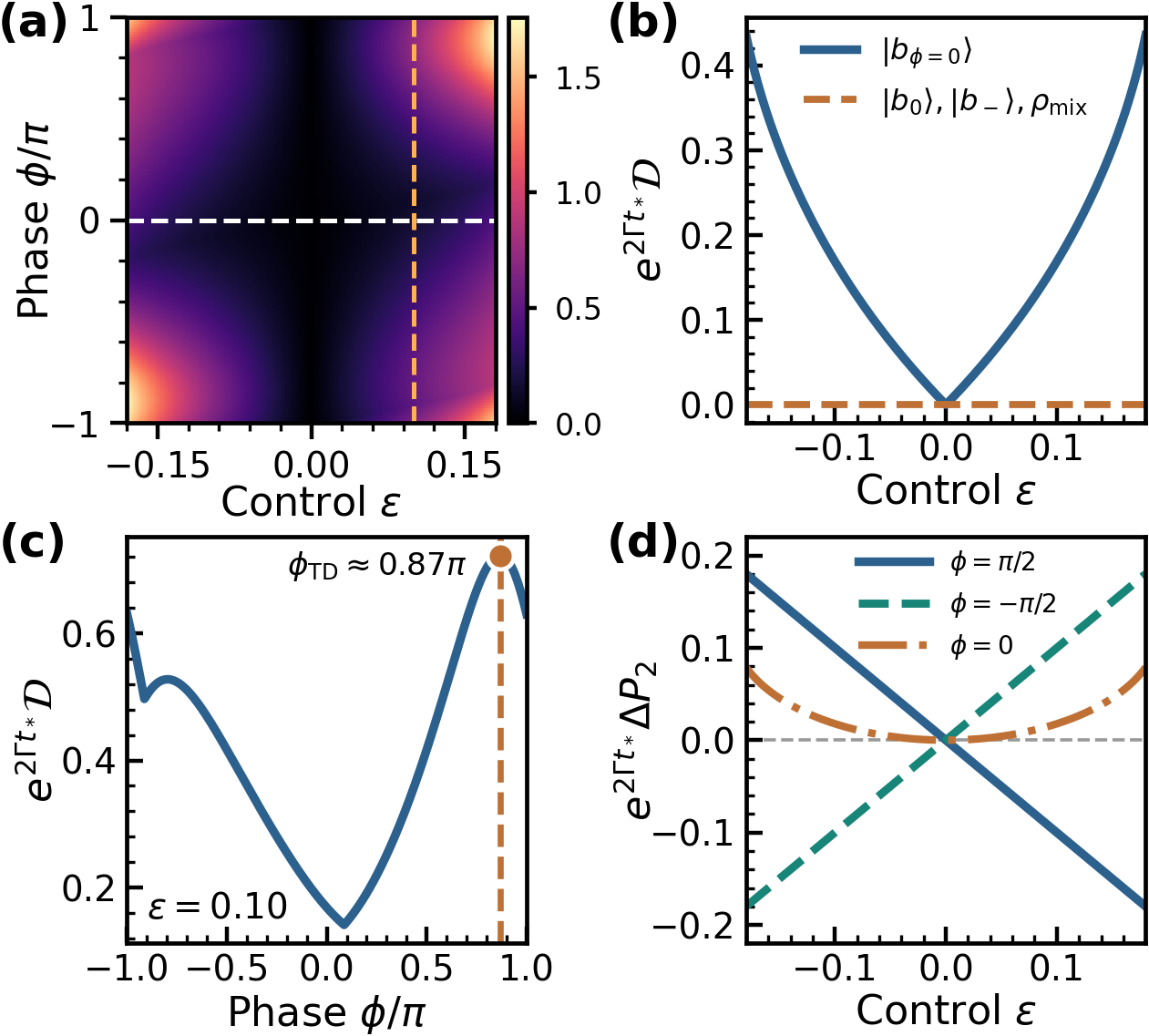} 
\caption{\textbf{Coherence-enabled recovery and population readout of hidden dynamical phases.}
(a) Coherence-resolved distinguishability landscape at $t_*=\sqrt3$ for the normalized superposition $|b_\phi\rangle$ defined in the text. The color scale shows the rescaled complete-output trace distance $e^{2\Gamma t_*}\mathcal D$. Dashed lines indicate the cuts $\phi=0$ and $\varepsilon=0.10$ in (b) and (c), respectively.
(b) Control dependence for the coherent preparation $|b_{\phi=0}\rangle$, compared with the individually blind preparations $|b_0\rangle$, $|b_-\rangle$, and their incoherent mixture $\rho_{\rm mix}$. The latter three remain exactly indistinguishable.
(c) Preparation-phase dependence at $\varepsilon=0.10$. The maximum within this superposition family occurs at $\phi_{\rm TD}\simeq0.87\pi$. 
(d) Rescaled population contrast $e^{2\Gamma t_*}\Delta P_2$ for a fixed measurement of level $|2\rangle$. The preparation phases $\phi=\pm\pi/2$ yield opposite, exactly linear responses $\mp\varepsilon$, whereas $\phi=0$ produces an even response. The common loss offset is $\Gamma=1.25$.
}
\label{fig:fig3}
\end{figure}

{\it Coherence-enabled recovery}---At $t_*=\sqrt3$, the individually blind preparations $|b_0\rangle \equiv |b_0(t_*)\rangle$ and $|b_-\rangle$ carry distinct phase factors for $\varepsilon\neq0$. Strikingly, coherent superposition of these preparations restores distinguishability between the two dynamics, while their incoherent mixture remains exactly blind. To demonstrate this, consider the incoherent mixture $\rho_{\rm mix} =(|b_0\rangle\langle b_0| +|b_-\rangle\langle b_-|)/2$ and the normalized coherent superposition $|b_\phi\rangle = (|b_0\rangle+e^{i\phi} |b_-\rangle) /\sqrt{2(1 +\cos\phi/\sqrt3)}$. At $\phi=0$, the coherent preparation $|b_{\rm coh}\rangle=|b_{\phi=0}\rangle$ restores distinguishability [Fig.~\ref{fig:fig3}(a,b)]: $\mathcal D_{\rm mix}(t_*)=0$, whereas $\mathcal D_{\rm coh}(t_*)>0$ for $\varepsilon\ne0$~\cite{SM}.  
Varying the preparation phase $\phi$ enables optimization of distinguishability within this superposition family [Fig.~\ref{fig:fig3}(a,c)].

More directly, the hidden dynamical phase can be read out through a fixed population measurement. Defining $\Delta P_2(\varepsilon,\phi) =P_2^{(\varepsilon)}(\phi)-P_2^{(0)}(\phi)$, where $P_2^{(\varepsilon)}(\phi) =|\langle 2|K_\varepsilon(t_*)|b_\phi\rangle|^2$, we obtain the exact response $e^{2\Gamma t_*}\Delta P_2 =[y(\varepsilon) \cos\phi/\sqrt{3} -\varepsilon\sin\phi]/ [1 +\cos\phi /\sqrt{3}]$. Remarkably, choosing $\phi=\pm\pi/2$ gives $e^{2\Gamma t_*}\Delta P_2=\mp\varepsilon$ exactly, whereas $\phi=0$ yields an even response [Fig.~\ref{fig:fig3}(d)]. Thus, preparation coherence converts an otherwise hidden relative phase into a directly measurable, sign-sensitive population signal.

More generally, coherent superposition of individually blind preparations with nonzero surviving amplitudes restores distinguishability when their relative-propagator phase factors differ. The general criterion, trace-distance bounds, phase optimization, and population-response derivation are given in Supplemental Material~\cite{SM}.

A complementary recovery strategy is preparation switching. At $t=2$, replacing the blind preparation $|b_{\mathrm B}\rangle$ with $|b_{\mathrm D}\rangle=|2\rangle$ enables a zero-background detection event upon projection onto $|3\rangle$, with unconditional probability $P_{\mathrm D}(\varepsilon)= 36 e^{-4\Gamma} \varepsilon^2 +O(\varepsilon^4)$. For the specified binary discrimination task, the optimal sample complexity is $n_{\mathrm{opt}}=\Theta(\varepsilon^{-2})$, even allowing collective measurements (see Supplemental Material~\cite{SM}).

{\it Krylov-space criteria for identifiability}---Although complete dynamical accessibility does not guarantee distinguishability at a fixed time, can it rule out persistent blindness? More generally, what physical access suffices for perturbation detection and Jordan-structure identification?
Consider $H(\varepsilon)=H_0 +\varepsilon V + O(\varepsilon^2)$, where $V=V^\dagger$, and denote by $\mathcal K_b =\operatorname{span} \{H_0^n |b\rangle:n \geq 0\}$ the Krylov subspace of the reference Hamiltonian $H_0$ generated by $|b\rangle$. 
First-order blindness means that the output derivative with respect to $\varepsilon$ vanishes at $\varepsilon=0$. It persists over a nonempty open time interval if and only if $V|_{\mathcal K_b}=cI|_{\mathcal K_b}$ for some $c\in\mathbb R$ (see Supplemental Material~\cite{SM}).
Physically, the perturbation must act as a common energy shift throughout the explored dynamical subspace, producing only an unobservable phase to first order. Otherwise, for a fixed preparation and perturbation direction, first-order blindness can occur only at isolated interrogation times.
For two time-independent Hamiltonians $H_0$ and $H_1$ with the same dissipation, exact blindness over an open time interval holds if and only if $(H_1-H_0)|_{\mathcal K_b}=cI|_{\mathcal K_b}$ for some real $c$ (see Supplemental Material~\cite{SM}).  
In particular, if the preparation is cyclic for either Hamiltonian, two fixed-dissipation Hamiltonians with different Jordan-block-size multisets can be exactly indistinguishable only at isolated interrogation times. Thus, complete dynamical accessibility does not preclude isolated blind times but guarantees that distinguishability between different Jordan structures can be recovered by varying the interrogation time~\cite{SM}. 

Beyond temporal access, preparation coherence provides a general resource for perturbation detection. In a $d$-dimensional system, the $d$ basis preparations $\{|j\rangle\}$ together with $|s\rangle =d^{-1/2} \sum_j|j\rangle$ guarantee that, for every non-scalar Hermitian first-order perturbation $V$, at least one preparation produces a nonzero complete-output response at sufficiently short, nonzero interrogation times. 
The leading short-time output derivative in the surviving sector is $-it[V,|b\rangle\langle b|]$: the basis preparations are insensitive to diagonal perturbations at leading order, whereas a single coherent superposition $|s\rangle$ makes relative energy shifts observable, leaving only a common energy shift undetectable at this order. For an isospectral family, the fixed eigenvalue sum implies $\operatorname{Tr}V=0$, making every nonzero Hermitian first-order control direction detectable. This coherence-enabled removal of blind perturbation directions is illustrated numerically in Supplemental Material~\cite{SM}.

Detection does not, however, imply identification of the Jordan partition. With $d^2$ informationally complete preparations and ideal full survivor-output tomography~\cite{Poyatos1997,Chuang1997}, the reconstructed process determines $K(t)=e^{-iHt}$ up to a global phase, and hence determines its Jordan-block sizes at any nonzero time. This identifies the block-size multiset, though not necessarily the original eigenvalue labels. Thus, even at a blind interrogation time, the Jordan-block sizes remain encoded in the full quantum process. The complete proofs and numerical demonstrations are given in Supplemental Material~\cite{SM}.

\begin{figure}
\centering
 \includegraphics[width=.99\columnwidth]{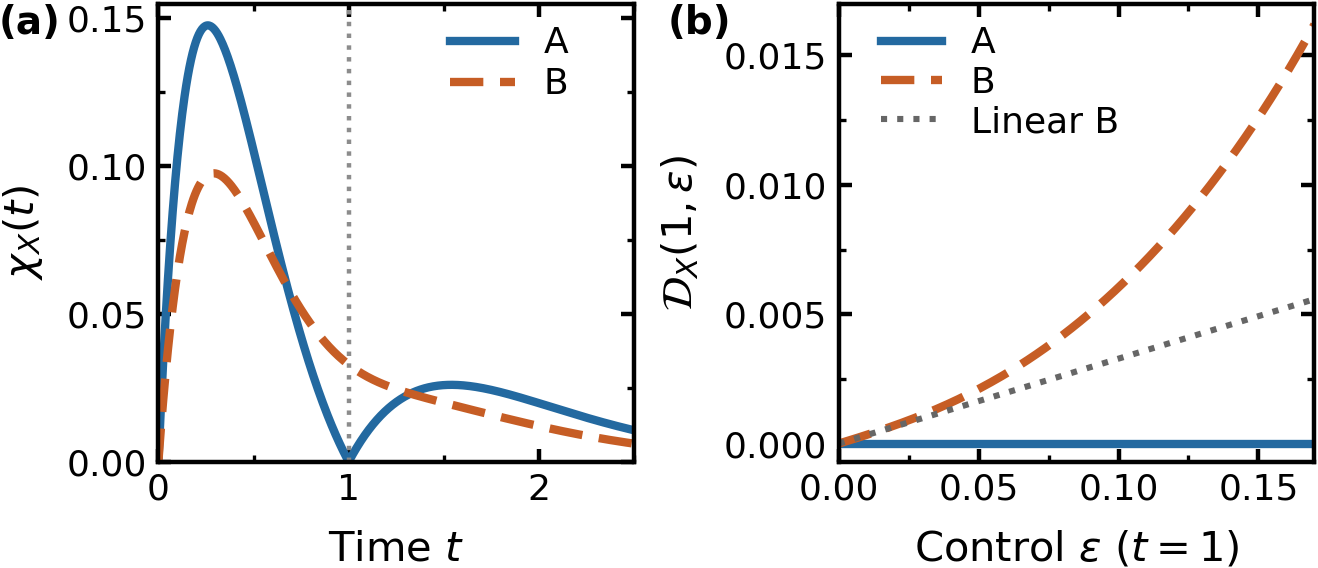} 
\caption{\textbf{Control-path dependence of operational
distinguishability.}
(a) Complete-output linear susceptibilities $\chi_X(t)$ for two isospectral coherent-control paths $X=A,B$, with identical dissipation and Jordan-partition transitions, probed using the common preparation $|b_{\rm c}\rangle =(-2i|1\rangle +i|2\rangle)/\sqrt5$. Path A is blind at $t=1$, whereas path B retains a finite response.
(b) Exact complete survival--loss trace distance $\mathcal D_X(1, \varepsilon)$ at $t=1$. Path A remains exactly blind throughout the displayed control interval, while path B yields a nonzero signal. The gray dotted line denotes the linear approximation $\chi_B(1)|\varepsilon|$. The common loss offset is $\Gamma=1.25$.
}
\label{fig:fig4}
\end{figure}

{\it Control-path-dependent identifiability}---Does operational distinguishability depend on the physical control path even when the spectral and Jordan-structural changes are identical? We compare two families, $N_A(\varepsilon)\equiv N(\varepsilon)$ and $N_B(\varepsilon)$, sharing the same reference Hamiltonian and fixed dissipation but differing in their coherent controls. Along both paths, $N_X^3(\varepsilon)=0$ ($X=A,B$), and the Jordan partition changes from $(2,1)$ to $(3)$ for sufficiently small nonzero $\varepsilon$ (see Supplemental Material~\cite{SM} for explicit constructions).

Using a common preparation for both paths, $|b_{\rm c}\rangle=(-2i|1\rangle +i|2\rangle)/\sqrt5$, we denote the corresponding complete-output trace distance along path $X$ by $\mathcal D_X(t,\varepsilon)$ and define the linear susceptibility $\chi_X(t) =\lim_{\varepsilon\to0} \mathcal D_X(t,\varepsilon) /|\varepsilon|$. Exact evaluation yields $\chi_A(t) =(3\sqrt6/5) e^{-2\Gamma t}|t(t-1)|$ and $\chi_B(t)=(2t/5) e^{-2\Gamma t} \sqrt{1+4(t-1)^2}$~\cite{SM}.
At $t=1$, path A is exactly blind throughout its analytic control branch, whereas $\chi_B(1) =2e^{-2\Gamma}/5>0$ [Fig.~\ref{fig:fig4}(a)]. The exact finite-control trace distances confirm that this distinction persists beyond linear response [Fig.~\ref{fig:fig4}(b)]. 
Strikingly, identical spectra, dissipation, and Jordan-partition transitions do not guarantee identical operational responses. The physical direction of coherent control provides independent information that is absent from the intrinsic Jordan structure: under the same preparation and interrogation time, one path can be exactly blind while another remains detectable.

{\it Physical realization and accessibility}---The model can be realized in the single-excitation sector of three coupled modes, with $|j\rangle$ denoting an excitation in mode $j$. The isospectral path in Eq.~(\ref{eq:N}) is generated by coordinating the onsite energies and the amplitudes and phases of the three pairwise coherent couplings through a single control parameter, while keeping all decay rates fixed. 
Programmable coupled-mode platforms, including superconducting circuits with parametrically controlled exchange interactions and synthetic coupling phases~\cite{Roushan2017,Han2020}, together with engineered dissipation~\cite{Shankar2013,Maurya2024}, offer possible implementations, provided the coherent controls and losses can be calibrated simultaneously.

The dissipative dynamics is described by the Lindblad operators $L_j=\sqrt{2d_j}|{\rm vac}\rangle\langle j|$, where $|{\rm vac}\rangle$ is the common vacuum and $(d_1,d_2,d_3)=(\Gamma+1,\Gamma,\Gamma-1)$. Together with the coherent Hamiltonian $C(\varepsilon)$, these decay channels yield the no-jump Hamiltonian $H(\varepsilon) =C(\varepsilon)-iD$. The surviving single-excitation sector and the vacuum population then constitute the complete survival--loss output, without postselection or resolution of individual decay channels. 

The predicted effects can be probed experimentally by switching preparations at $t=2$, comparing blind inputs with their coherent superpositions at $t_*=\sqrt3$, and testing both control paths with the same preparation at $t=1$. In particular, preparation switching and phase-controlled superposition yield zero-background and sign-sensitive population signals, respectively, without requiring full state tomography for detection. 
Uniform additional loss preserves exact blindness, whereas differential-loss errors and pure dephasing generally lift it. The effects of control imperfections, noise, and finite sampling, together with implementation details, are analyzed in Supplemental Material~\cite{SM}.

{\it Discussion}---Dynamical accessibility does not guarantee physical identifiability: destructive interference can conceal an isospectral Jordan transition in the complete survival--loss output at a finite time, despite full access to the emerging Jordan chain.

Such blindness is protocol dependent, rather than an intrinsic obstruction to recovering the Jordan structure. Coherent preparations can expose hidden relative phases, while our general Krylov-space criteria distinguish isolated blind times from persistent blindness and identify sufficient resources for detection and Jordan-structure reconstruction. Furthermore, different coherent-control paths can make the same Jordan transition either invisible or detectable under an identical probing protocol. 
These results establish physical identifiability as a distinct, protocol-dependent aspect of exceptional degeneracies, governed not only by their intrinsic Jordan structures but also by the available preparations, interrogation times, coherent controls, and measurements.

{\it Acknowledgments}---This work was supported by the National Key Research and Development Program of China (Grant No.~2025YFE0217400) and the National Natural Science Foundation of China (Grant No.~12674619).

{\it Data availability}---The data supporting the findings of this study are provided in the main text and Supplemental Material. Additional numerical data are available from the corresponding authors upon reasonable request.


\begin{thebibliography}{99}

\bibitem{Kato}
T. Kato,
\textit{Perturbation Theory for Linear Operators},
2nd ed. (Springer-Verlag, Berlin, 1976).

\bibitem{AshidaGongUeda20AdvPhys}
Y. Ashida, Z. Gong, and M. Ueda,
{\it Non-Hermitian Physics}, 
\href{https://doi.org/10.1080/00018732.2021.1876991}{Adv. Phys. {\bf 69}, 249 (2020)}.

\bibitem{Bender2024}
C. M. Bender and D. W. Hook,
${\cal PT}$-symmetric quantum mechanics,
\href{https://doi.org/10.1103/RevModPhys.96.045002}{Rev. Mod. Phys. 96, 045002 (2024)}.

\bibitem{Heiss2012}
W. D. Heiss,
The physics of exceptional points,
\href{https://doi.org/10.1088/1751-8113/45/44/444016}{J. Phys. A: Math. Theor. \textbf{45}, 444016 (2012)}.

\bibitem{ElGanainy2018}
R. El-Ganainy, K. G. Makris, M. Khajavikhan, Z. H. Musslimani, S. Rotter, and D. N. Christodoulides,
Non-Hermitian physics and PT symmetry,
\href{https://doi.org/10.1038/nphys4323}{Nat. Phys. \textbf{14}, 11 (2018)}.

\bibitem{Miri2019}
M.-A. Miri and A. Al\`u,
Exceptional points in optics and photonics,
\href{https://doi.org/10.1126/science.aar7709}{Science \textbf{363}, eaar7709 (2019)}.

\bibitem{Bergholtz2021}
E. J. Bergholtz, J. C. Budich, and F. K. Kunst,
Exceptional topology of non-Hermitian systems,
\href{https://doi.org/10.1103/RevModPhys.93.015005}{Rev. Mod. Phys. \textbf{93}, 015005 (2021)}.

\bibitem{Xue2026}
P. Xue, 
Essay: Topological phases and exceptional points in non-Hermitian systems,
\href{https://doi.org/10.1103/ll76-j2l5}{Phys. Rev. Lett. {\bf 136}, 170001 (2026)}.

\bibitem{Ding2022}
K. Ding, C. Fang, and G. Ma,
Non-Hermitian topology and exceptional-point geometries,
\href{https://doi.org/10.1038/s42254-022-00516-5}{Nat. Rev. Phys. \textbf{4}, 745 (2022)}.

\bibitem{Dembowski2001}
C. Dembowski, H.-D. Gr\"af, H. L. Harney, A. Heine, W. D. Heiss, H. Rehfeld, and A. Richter,
Experimental observation of the topological structure of exceptional points,
\href{https://doi.org/10.1103/PhysRevLett.86.787}{Phys. Rev. Lett. \textbf{86}, 787 (2001)}.

\bibitem{Dembowski2004}
C. Dembowski, B. Dietz, H.-D. Gr\"af, H. L. Harney, A. Heine, W. D. Heiss, and A. Richter,
Encircling an exceptional point,
\href{https://doi.org/10.1103/PhysRevE.69.056216}{Phys. Rev. E \textbf{69}, 056216 (2004)}.

\bibitem{Doppler2016}
J. Doppler, A. A. Mailybaev, J. B\"ohm, U. Kuhl, A. Girschik, F. Libisch, T. J. Milburn, P. Rabl, N. Moiseyev, and S. Rotter,
Dynamically encircling an exceptional point for asymmetric mode switching,
\href{https://doi.org/10.1038/nature18605}{Nature \textbf{537}, 76 (2016)}.

\bibitem{XuHarris16nature} 
H. Xu, D. Mason, L. Jiang, and J. G. E. Harris, 
Topological energy transfer in an optomechanical system with exceptional points, 
\href{https://doi.org/10.1038/nature18604}{Nature {\bf 537}, 80 (2016)}.

\bibitem{Li2023}
Z.-Z. Li, W. Chen, M. Abbasi, K. W. Murch, and K. B. Whaley,
Speeding up entanglement generation by proximity to higher-order exceptional points,
\href{https://doi.org/10.1103/PhysRevLett.131.100202}{Phys. Rev. Lett. \textbf{131}, 100202 (2023)}.

\bibitem{LiXue2026QuanAdv}
Z.-Z. Li and X. Xue,
Quantum advantage by designing beneficial dissipation from lossless dynamics,
\href{https://arxiv.org/abs/2610.38183}{arXiv:2610.38183 (2026)}. 

\bibitem{Hodaei2017}
H. Hodaei, A. U. Hassan, S. Wittek, H. Garcia-Gracia, R. El-Ganainy, D. N. Christodoulides, and M. Khajavikhan,
Enhanced sensitivity at higher-order exceptional points,
\href{https://doi.org/10.1038/nature23280}{Nature \textbf{548}, 187 (2017)}.

\bibitem{Chen2017}
W. Chen, Ş. K. \"{O}zdemir, G. Zhao, J. Wiersig, and L. Yang, 
Exceptional points enhance sensing in an optical microcavity, 
\href{https://doi.org/10.1038/nature23281}{Nature \textbf{548}, 192 (2017)}. 

\bibitem{Zhong2018}
Q. Zhong, M. Khajavikhan, D. N. Christodoulides,
and R. El-Ganainy,
Winding around non-Hermitian singularities,
\href{https://doi.org/10.1038/s41467-018-07105-0}{Nat. Commun. \textbf{9}, 4808 (2018)}.

\bibitem{NaghilooMurch19nphys} 
M. Naghiloo, M. Abbasi, Y. N. Joglekar, and K. W. Murch, 
Quantum state tomography across the exceptional point in a single dissipative qubit,
\href{https://doi.org/10.1038/s41567-019-0652-z}{Nat. Phys. {\bf 15}, 1232 (2019)}. 

\bibitem{WuRongDu19science}
Y. Wu, W. Liu, J. Geng, X. Song, X. Ye, C. K. Duan, X. Rong, and J. Du, 
Observation of parity-time symmetry breaking in a single-spin system,
\href{https://doi.org/10.1126/science.aaw8205}{Science {\bf 364}, 878 (2019)}.

\bibitem{DingZhang21prl}
L. Ding, K. Shi, Q. Zhang, D. Shen, X. Zhang, and W. Zhang, 
Experimental determination of ${\cal PT}$-symmetric exceptional points in a single trapped ion,
\href{https://doi.org/10.1103/PhysRevLett.126.083604}{Phys. Rev. Lett. {\bf 126}, 083604 (2021)}.

\bibitem{WangJingChen21pra}
W. C. Wang, Y. L. Zhou, H. L. Zhang, J. Zhang, M. C. Zhang, Y. Xie, C. W. Wu, T. Chen, B. Q. Ou, W. Wu, H. Jing, and P. X. Chen, 
Observation of ${\cal PT}$-symmetric quantum coherence in a single-ion system,
\href{https://doi.org/10.1103/PhysRevA.103.L020201}{Phys. Rev. A {\bf 103}, L020201 (2021)}. 

\bibitem{Ryu2024}
J.-W. Ryu, J.-H. Han, C.-H. Yi, M. J. Park, and H. C. Park,
Exceptional classifications of non-Hermitian systems,
\href{https://doi.org/10.1038/s42005-024-01595-9}{Commun. Phys. \textbf{7}, 109 (2024)}.

\bibitem{LiWhaley2024}
Z.-Z. Li and K. B. Whaley,
Enhancement of vibrationally assisted energy transfer by proximity to exceptional points, probed by fluorescence-detected vibrational spectroscopy,
\href{https://doi.org/10.1103/PhysRevResearch.6.023149}{Phys. Rev. Research \textbf{6}, 023149 (2024)}.

\bibitem{LiXue2026}
Z.-Z. Li and X. Xue,
Synchronizing spectral and interference criticalities at an exceptional point,
\href{https://arxiv.org/abs/2609.30311}{arXiv:2609.30311 (2026)}.

\bibitem{LiWhaley2026}
Z.-Z. Li and K. B. Whaley,
Beyond exceptional-point order 
(unpublished). 

\bibitem{LiMa2026}
Z. Li, R. Cai, X. Wang, K. Shimomura, C. Lu, Z. Yang, M. Sato, and G. Ma,
Exceptional deficiency of non-Hermitian systems,   
\href{https://doi.org/10.1038/s41567-026-03259-7}{Nat. Phys. {\bf 22}, 962 (2026)}.

\bibitem{Shiralieva2026}
A. Shiralieva, G. A. Starkov, and B. Trauzettel,
Multiblock exceptional points in open quantum systems,
\href{https://doi.org/10.1103/gtw6-yn95}{Phys. Rev. A \textbf{113}, 052202 (2026)}.

\bibitem{Brandao2026}
P. A. Brand\~ao,
Operational signatures of exceptional quantum channels,
\href{https://doi.org/10.48550/arXiv.2610.13622}{arXiv:2610.13622 (2026)}.

\bibitem{Ghosh2026}
A. Ghosh and M. Bhattacharya,
Krylov tomography and finite-uncertainty certification of exceptional-point dynamics,
\href{https://arxiv.org/abs/2608.19761}{arXiv:2608.19761 (2026)}.

\bibitem{Lau2018}
H.-K. Lau and A. A. Clerk,
Fundamental limits and non-reciprocal approaches in non-Hermitian quantum sensing,
\href{https://doi.org/10.1038/s41467-018-06477-7}{Nat. Commun. \textbf{9}, 4320 (2018)}. 

\bibitem{Burgarth2012}
D. Burgarth and K. Yuasa,
Quantum system identification,
\href{https://doi.org/10.1103/PhysRevLett.108.080502}{Phys. Rev. Lett. \textbf{108}, 080502 (2012)}.

\bibitem{Burgarth2014}
D. Burgarth and K. Yuasa,
Identifiability of open quantum systems,
\href{https://doi.org/10.1103/PhysRevA.89.030302}{Phys. Rev. A \textbf{89}, 030302(R) (2014)}.

\bibitem{SM}
See Supplemental Material at [URL will be inserted by publisher] for the exact isospectral constructions, finite-time blindness, timing and calibration-error sensitivity, statistical detection costs, coherence-enabled recovery, control-path-dependent susceptibilities, proofs of the general physical-access and reconstruction results, and details of physical implementation and noise robustness, including citations of Refs.~\cite{Fuchs1999,Chernoff1952}.

\bibitem{Helstrom1976}
C. W. Helstrom,
\textit{Quantum Detection and Estimation Theory}
(Academic Press, New York, 1976).

\bibitem{Poyatos1997}
J. F. Poyatos, J. I. Cirac, and P. Zoller,
Complete characterization of a quantum process: The two-bit quantum gate,
\href{https://doi.org/10.1103/PhysRevLett.78.390}{Phys. Rev. Lett. \textbf{78}, 390 (1997)}.

\bibitem{Chuang1997}
I. L. Chuang and M. A. Nielsen,
Prescription for experimental determination of the dynamics of a quantum black box,
\href{https://doi.org/10.1080/09500349708231894}{J. Mod. Opt. \textbf{44}, 2455 (1997)}.

\bibitem{Roushan2017}
P. Roushan, C. Neill, A. Megrant, Y. Chen, R. Babbush, R. Barends, B. Campbell, Z. Chen, B. Chiaro, A. Dunsworth, A. Fowler, E. Jeffrey, J. Kelly, E. Lucero, J. Mutus, P. J. J. O’Malley, M. Neeley, C. Quintana, D. Sank, A. Vainsencher, J. Wenner, T. White, E. Kapit, H. Neven, and J. Martinis, 
Chiral ground-state currents of interacting photons in a synthetic magnetic field,
\href{https://doi.org/10.1038/nphys3930}{Nat. Phys. {\bf 13}, 146 (2017)}. 

\bibitem{Han2020}
X. Y. Han, T. Q. Cai, X. G. Li, Y. K. Wu, Y. W. Ma, Y. L. Ma, J. H. Wang, H. Y. Zhang, Y. P. Song, and L. M. Duan,
Error analysis in suppression of unwanted qubit interactions for a parametric gate in a tunable superconducting circuit,
\href{https://doi.org/10.1103/PhysRevA.102.022619}{Phys. Rev. A {\bf 102}, 022619 (2020)}.

\bibitem{Shankar2013}
S. Shankar, M. Hatridge, Z. Leghtas, K. M. Sliwa, A. Narla, U. Vool, S. M. Girvin, L. Frunzio, M. Mirrahimi, and M. H. Devoret, 
Autonomously stabilized entanglement between two superconducting quantum bits,
\href{https://doi.org/10.1038/nature12802}{Nature {\bf 504}, 419 (2013)}.

\bibitem{Maurya2024}
V. Maurya, H. Zhang, D. Kowsari, A. Kuo, D. M. Hartsell, C. Miyamoto, J. Liu, S. Shanto, E. Vlachos, A. Zarassi, K. W. Murch, and E. M. Levenson-Falk, 
On-Demand Driven Dissipation for Cavity Reset and Cooling,
\href{https://doi.org/10.1103/PRXQuantum.5.020321}{PRX Quantum \textbf{5}, 020321 (2024)}.

\bibitem{Fuchs1999}
C. A. Fuchs and J. van de Graaf,
Cryptographic distinguishability measures for quantum-mechanical states,
\href{https://doi.org/10.1109/18.761271}{IEEE Trans. Inf. Theory \textbf{45}, 1216 (1999)}.

\bibitem{Chernoff1952}
H. Chernoff, 
A Measure of Asymptotic Efficiency for Tests of a Hypothesis Based on the Sum of Observations,
\href{https://doi.org/10.1214/aoms/1177729330}{Ann. Math. Statist. {\bf 23}, 493 (1952)}.



\end{thebibliography}
\end{document}